# Fairy chimneys in Peru

Amelia Carolina Sparavigna
Dipartimento di Fisica,
Politecnico di Torino, Torino, Italy

Erosion creates beautiful landscapes. In most cases, it is just the local population to know them. Google Maps can be helpful for locating the places, for their studies and eventual projects of preservation. An interesting example is given by a landscape of fairy chimneys in Peru, near Pampachiri and San Pedro de Larcay. It is remarkable the fact that some of them have been adapted as dwelling places.

The erosion of water and wind on the rocks is able to create beautiful landscapes. Some places are well known and have been declared natural heritage sites by UNESCO; some others are in desert and hostile regions, hard to visit. In most cases, just the local population knows these landscapes. In this paper, we will see that the simple action of some users of World Wide Web services, such as the Google Maps, in uploading their pictures, allows the discovery, study and perhaps future preservation of some of them. We will see in particular the case of some fairy chimneys in Peru.
Before talking about this Peruvian landscape, let us discuss briefly the nature of these rocky structures, tall and thin cusps of rock protruding from the land. They are known in several manners [1]. These pinnacles are considered as "tent rocks", "fairy chimneys" or simply "earth pyramids", according to their shapes. Another name is "hoodoo": at first sight, this name seems to be derived from "hood" because of the structure looking like a sort of "dwarf hat", but probably its etymology is different [2]. Hoodoo is the common term used to describe the rock chimneys found in the western United States and Canada.
The fairy chimneys consist of a relatively soft rock. Some of them have on the top a harder stone, less easy to erode: the result is a geological structure resembling that of a chimney. These structures typically arise within sedimentary rocks or volcanic formations. The height of these chimneys can be of tenths of meters. Their shapes are affected by the existence of different and alternate layers of materials having different strength. In some regions, there are the freezing of water and the gravity that are creating these structures. The process is known as "frost wedging". The water, that percolates in the cracks of rocks, freezes and then expands, acting as a wedge and breaking the rocks apart [1,3]. This is the same action chiseling the landscape of Dolomites, the mountain range located in northeastern Italy [4]. During the night, when the temperature goes below the freezing point, the water into the fractures of rock turns into ice. The corresponding expansion of volume increases the distance between the sides of fractures. During the day, the sun warms the rocks and water melts. So separated from the bulk, some parts of the rock fall for gravity. In the case that the fairy chimneys are made by tuff rocks from volcanic eruptions, the erosion is due to wind and rain.
In Italy, there are several areas with gullies (calanchi) and pinnacles, chiseled by water and wind. Well-known are the Calanchi di Volterra in Tuscany, but several others interesting places are in Abruzzo [5].
Among the best-known landscapes having fairy chimneys, there is that of Cappadocia [6]. Besides the importance of this geophysical area, the region is quite interesting because Cappadocians carved their homes into the soft rock (see Fig.1). During the medieval era, this area becomes a refuge for Byzantine Christians. The people established monastic settlements and churches inside the pinnacles. According to [6], the Göreme Open-Air Museum in Cappadocia has the best-preserved collection of chapels and houses, most dating about the 11th Century [7]. The life in Cappadocia was even more complex, because, due to the persecution, the local Christians often had to hide themselves. It seems that, alarmed by the hoof beats [6], they could abandon the caves in the pinnacles to find a refuge in the underground. Beneath the ground of Cappadocia, archaeologists found a network of subterranean villages, the largest discovered is almost 10 levels deep, with narrow passages among them [6-9].

Not only Cappadocia has some houses inside the natural chimneys. We can find them also in Peru, in a location near Pampachiri and San Pedro de Larcay. This is a region having several interesting places for geophysical and archaeological researches. Near San Pedro in fact, there is the archaeological site of Wallpa Wiri, having structures of Incas age [10]. Moreover, this region possesses a large structured system of carved stones, used as landmarks for agricultural purposes, system created in the Late Horizon period of Peruvian prehistory [11]. Analyzing this landscape within its agricultural and social context [11[, the researcher provided evidence of water distribution and management of the irrigation cycles by the Incas; that is, that there was an administration and management of the local agricultural processes by the central government. A survey of the region with Google Maps reveals the presence, at south-west of San Pedro, coordinates -14.178624,-73.592713, of several "qochas", ponds with a diameters of about 100 meters (see Fig.2). Linked together by a network of canals, qochas form a system of water and soil management [12]. This system is probably of pre-Incaic origin.

Unfortunately, the satellite map corresponding to this area has a limited resolution. However, Google Maps has an interesting feature: it is possible to drag the icon of the street view on the map and see if there are photos of the landscape uploaded by users. Doing this dragging, the location of the pictures appears as blue dots on the map (see Fig.3). In this manner, anybody using the map can have some information on the landscape corresponding to the specific location. There are no photos of the qochas of San Pedro, but many pictures of nearby locations, uploaded by Max Altamirano Moler [13]. Besides being very beautiful, the pictures display the existence of a forest of fairy chimneys. The location is given in the map of Fig.3.

It would be interesting to have some reports on the geophysical investigations of this region (probably documents in English or Spanish exist, but and are not available on the Web). To the author's knowledge, the pictures by Altamirano are the only existing detailed documentation on fairy chimneys in Peru [14], free on the Web. In fact, among the pictures that Altamirano has collected on a site (see Fig.4 and directly at [13]), several images are showing that some dwelling places have been obtained under or inside the fairy chimneys. It seems, as far as it is possible to gain from his pictures that a supporting structures of stones had built to reinforce the chimney (Fig.5). If not already existing, only a local survey and analysis can tell us how many and how old these structures are. In my opinion, a model of the thermodynamic behavior of these structures and of those in Cappadocia, with a lumped elements approach as in Ref.14, could be interesting to see if there are some thermal benefits in using them as houses.

In conclusion, the paper shows that a fairy chimneys landscape exists in Peru, where some structures seem used as dwelling places. Moreover, the documentation of these remarkable structures is due to the activity of a Google Maps user, demonstrating that each user can help in the process of spreading the knowledge of existing geological and cultural landscapes.

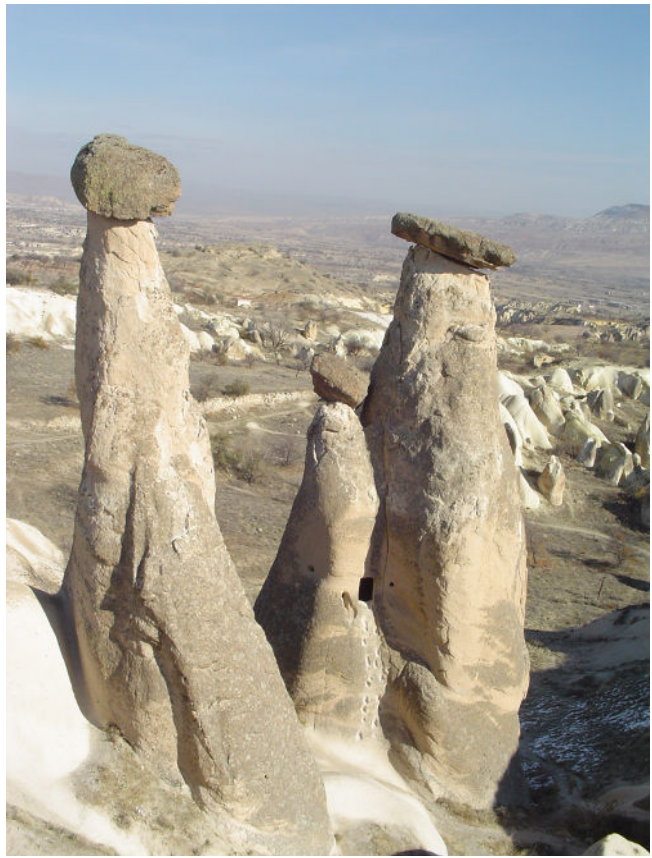

Fig.1. Fairy chimneys in Cappadocia (picture by Zeynel Cebeci, of a site at Ürgüp - Nevşehir, Turkey).

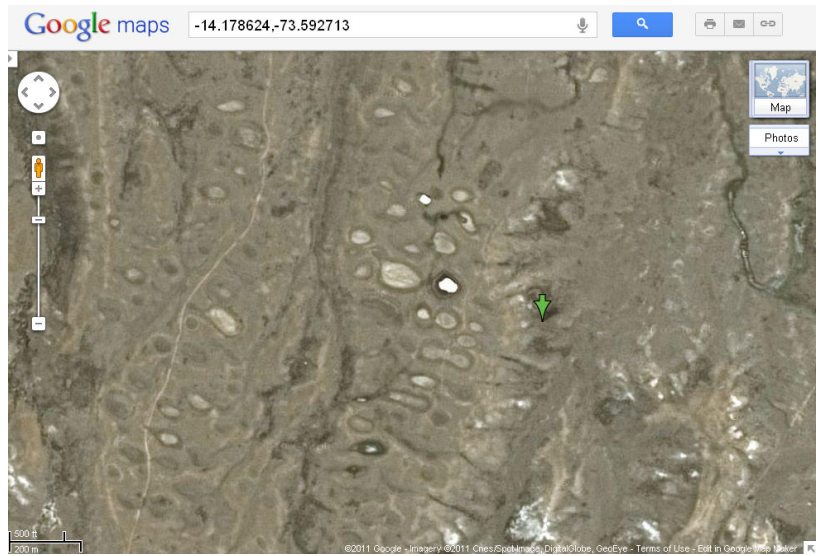

Fig.2. Some qochas near San Pedro. The average size is about 100 meters.

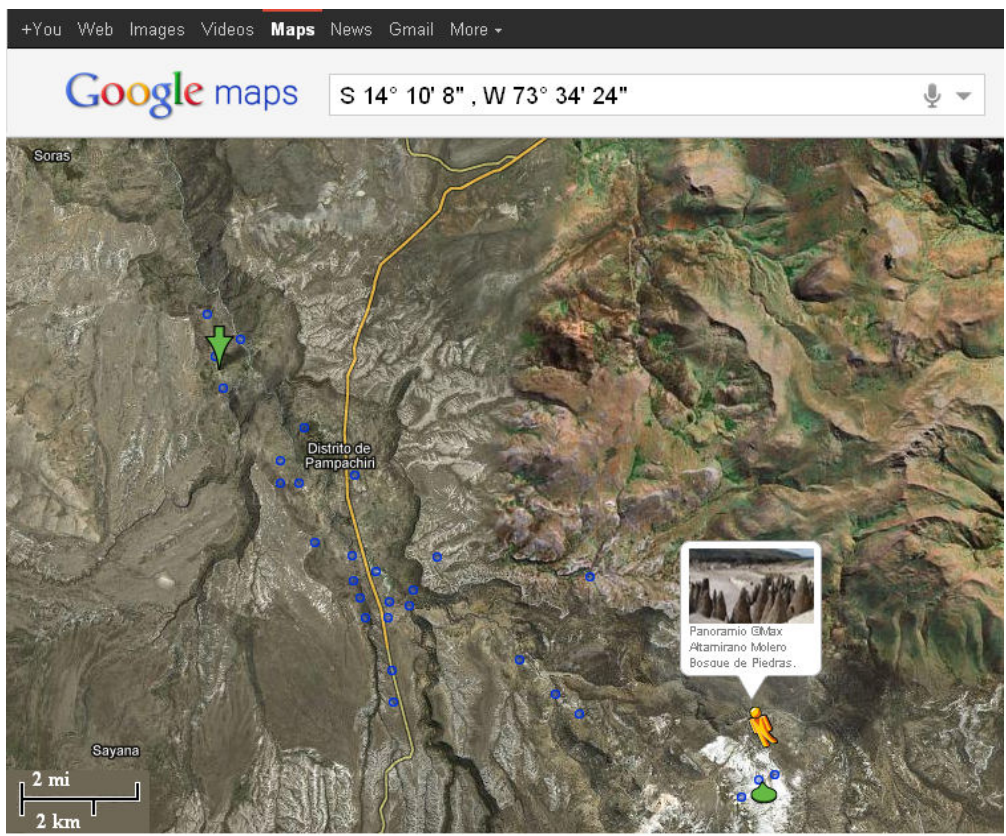

Fig.3. The position of San Pedro de Larcay, Peru, is given by the green arrow. The location of the fairy chimneys, near Pampachiri, is given by the blue dots corresponding to the photos.

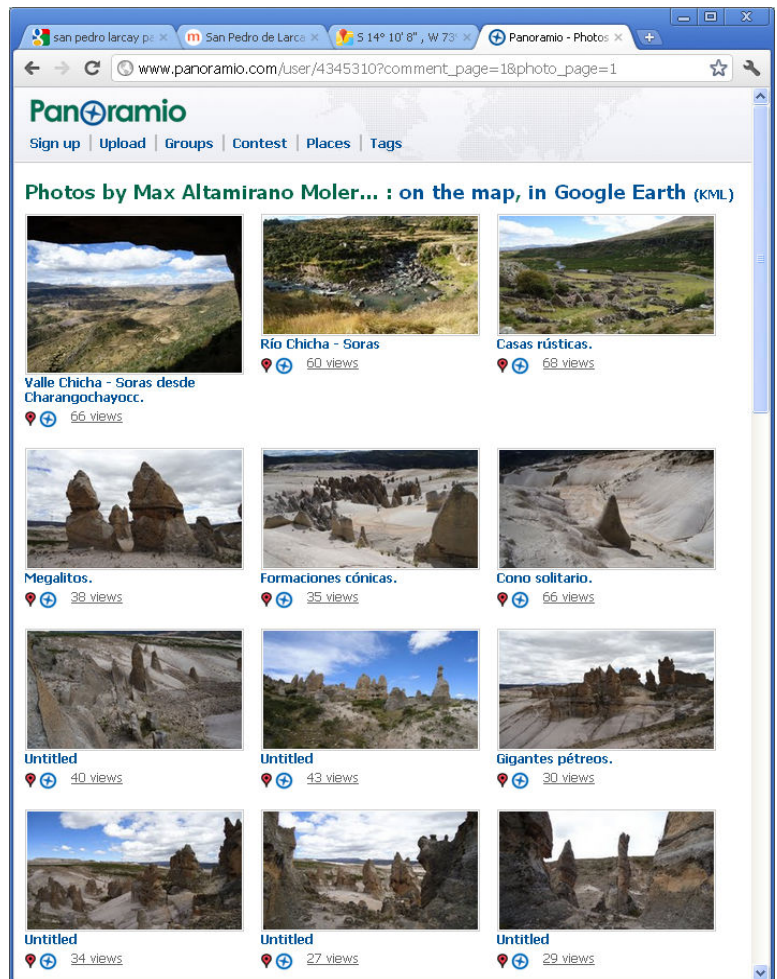

Fig.4. The image shows a part of the very interesting set of images by Max Altamirano Moler. To see them, the reader can visit [13].

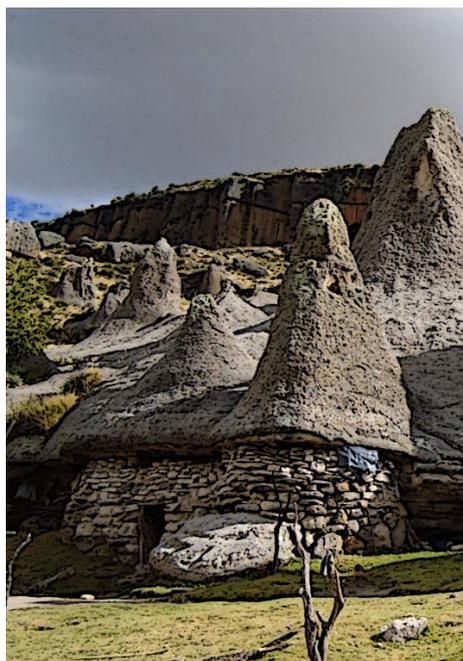

Fig.5. The image is adapted from a picture by Max Altamirano Moler, to show the importance of the pictures he collected. Note how the fairy chimneys had been transformed in a dwelling place.